\documentclass[twocolumn,superscriptaddress,prl]{revtex4}
\usepackage{graphicx}
\usepackage{epsfig}
\usepackage{color}
\usepackage{amsmath,amssymb,subfigure,psfrag}
\def\U#1{{\rm #1}} 
\def\u#1{_{\rm #1}}

\newcommand{\ket}[1]{| #1 \rangle}

\newcommand{\expect}[1]{\left\langle #1 \right\rangle} 
\newcommand{\dagg}[1]{#1 ^\dagger}

\def\H{{\rm H}}
\def\V{{\rm V}}

\begin{document}
\title{
Wide-band quantum interface for 
visible-to-telecommunication wavelength conversion 
}

\author{Rikizo Ikuta}
\affiliation{Graduate School of Engineering Science, Osaka University,
1-3 Machikaneyama, Toyonaka, Osaka 560-8531, Japan}
\author{Yoshiaki Kusaka}
\affiliation{Graduate School of Engineering Science, Osaka University,
1-3 Machikaneyama, Toyonaka, Osaka 560-8531, Japan}
\author{Tsuyoshi Kitano}
\affiliation{Graduate School of Engineering Science, Osaka University,
1-3 Machikaneyama, Toyonaka, Osaka 560-8531, Japan}
\author{Hiroshi Kato}
\affiliation{Graduate School of Engineering Science, Osaka University,
1-3 Machikaneyama, Toyonaka, Osaka 560-8531, Japan}
\author{Takashi Yamamoto}
\affiliation{Graduate School of Engineering Science, Osaka University,
1-3 Machikaneyama, Toyonaka, Osaka 560-8531, Japan}
\author{Masato Koashi}
\affiliation{
Photon Science Center, 
The University of Tokyo, 
2-11-16 Yayoi, 
Bunkyo-ku, Tokyo 113-8656, Japan} 
\author{Nobuyuki Imoto}
\affiliation{Graduate School of Engineering Science, Osaka University,
1-3 Machikaneyama, Toyonaka, Osaka 560-8531, Japan}

\begin{abstract}
We perform the first demonstration 
of a quantum interface for frequency down-conversion 
from visible to telecommunication bands by using a nonlinear crystal. 
This interface has a potential to work over wide bandwidths, 
leading to a high-speed interface of frequency conversion. 
We achieve the conversion of 
a pico-second visible photon at 780 nm to a 1522-nm photon, 
and observe that the conversion process 
retain entanglement 
between the down-converted photon and another photon. 
\end{abstract}
\maketitle

Photons play an important role for transmitting 
a quantum state among distantly located physical systems, 
such as atoms, ions, and 
solid-state systems~\cite{E1, E2, E4, E5, E6}, 
which are used 
as memories and/or processors 
of quantum communication~\cite{R1, R2} and 
information processing\cite{CZ1}. 
So far, experimental demonstrations 
aiming at such tasks 
have been actively done with photons 
in visible wavelengths~\cite{Ex1, Ex2}. 
However, when we look 
at long-distance quantum communication~\cite{communication}, 
a near-infrared wavelength 
within telecommunication bands is vital 
for efficient optical fiber communications. 
This optical frequency mismatch can be filled 
by frequency conversion of a photon 
while preserving its quantum state~\cite{conv}. 
A pioneer work of such a quantum interface 
for the optical frequency conversion has been done 
by using a second-order nonlinearity 
in a periodically-poled LiNbO$\u{3}$~(PPLN) crystal~\cite{PPLN1}, 
in which a photon of a telecom wavelength 
was up-converted to a visible one. 
Such a frequency up-conversion was also 
actively studied for efficient detection of photons 
in telecommunication bands~\cite{Langrock}. 
On the other hand, quantum interfaces 
for frequency down-conversion 
from visible 
to telecommunication bands have recently been demonstrated 
by using a third-order nonlinearity 
of a dense atomic cloud~\cite{four1, four2}, 
which uses two pump lights 
MHz-detuned from resonant frequencies of the atom. 
In these demonstrations, 
choice of the target wavelength 
is severely limited 
to the vicinity of the resonant frequency of the atom. 
Such a restriction on frequency selections can be relaxed 
by using a bulk nonlinear crystal. 
So far, aiming at such a versatile quantum interface, 
demonstrations of 
difference-frequency generation~(DFG) 
in nonlinear optical crystals 
with a strong pump and a weak coherent light seed 
have been performed~\cite{DFG1, DFG2}. 

In this Letter, we present the first demonstration 
of DFG-based frequency down-conversion of non-classical light 
by a nonlinear optical crystal 
from a visible wavelength at 780 nm 
to a telecommunication wavelength at 1522 nm. 
Using a waveguided PPLN crystal 
with a wide acceptable bandwidth 
as the second-order nonlinear optical material, 
we have observed 
sub-Poissonian photon statistics 
of the light at 1522 nm 
after the frequency down-conversion 
of a pico-second single-photon pulse at 780 nm. 
Then, we have demonstrated the frequency down-conversion of 
one half of a 780-nm entangled photon pair, 
and observed the entanglement 
between 
the down-converted photon at 1522 nm and the other half at 780 nm. 

The quantum theory of frequency conversion 
of single-mode pulsed light 
using a second-order nonlinear optical interaction is as follows. 
When the pump light at angular frequency $\omega\u{p}$ 
is sufficiently strong, 
the quantum dynamics of a signal mode at angular frequency $\omega\u{s}$ 
and a converted mode at angular frequency $\omega\u{c}$ 
satisfying $\omega\u{c}=\omega\u{s} - \omega\u{p}$ 
is described by the following Hamiltonian: 
\begin{eqnarray}
{\hat{H}}=i\hbar (\xi^*\dagg{\hat{a}}\u{c}\hat{a}\u{s}-\xi\dagg{\hat{a}}\u{s}\hat{a}\u{c})\ , 
\label{Hemiltonian}
\end{eqnarray}
where 
$\hat{a}\u{s}$ and $\hat{a}\u{c}$ are annihilation operators 
of the signal mode and the converted mode, respectively. 
A coupling constant $\xi = |\xi|e^{i\phi}$ of the nonlinear optical medium 
is proportional to the complex amplitude of the classical pump light, 
where $\phi$ is the phase of the pump light. 
Using the Heisenberg representation 
$\hat{a}\u{s(c)}(t)\equiv\dagg{\hat{U}}\hat{a}\u{s(c)}\hat{U}$ 
with 
$\hat{U}\equiv \exp(-i\hat{H}t/\hbar)$, 
annihilation operators $\hat{a}\u{s, out}$ and $\hat{a}\u{c, out}$ 
of the signal and converted modes coming out from the nonlinear medium 
are written as 
\begin{eqnarray}
\hat{a}\u{s,out}=\hat{a}\u{s}(\tau)&=
\cos(|\xi| \tau)\ \hat{a}\u{s} - e^{i\phi}\sin(|\xi| \tau)\ \hat{a}\u{c}
\label{as}
\end{eqnarray}
and 
\begin{eqnarray}
\hat{a}\u{c,out}=\hat{a}\u{c}(\tau)&=
 e^{-i\phi}\sin(|\xi| \tau)\ \hat{a}\u{s} +\cos(|\xi| \tau)\ \hat{a}\u{c}, 
\label{ac}
\end{eqnarray}
where $\tau$ is the traveling time of the pulses 
through the nonlinear medium. 
The process of the frequency conversion 
described in equations~(\ref{as}) and (\ref{ac}) can be regarded 
as a beamsplitter 
in the frequency degree of freedom 
with the transmittance of $\cos^2(|\xi| \tau)$. 
A condition $|\xi| \tau=\pi/2$ 
for the complete conversion is achieved 
by adjusting the amplitude of the pump light and 
the length of the nonlinear medium. 

\begin{figure}
 \begin{center}
 \scalebox{1}{\includegraphics{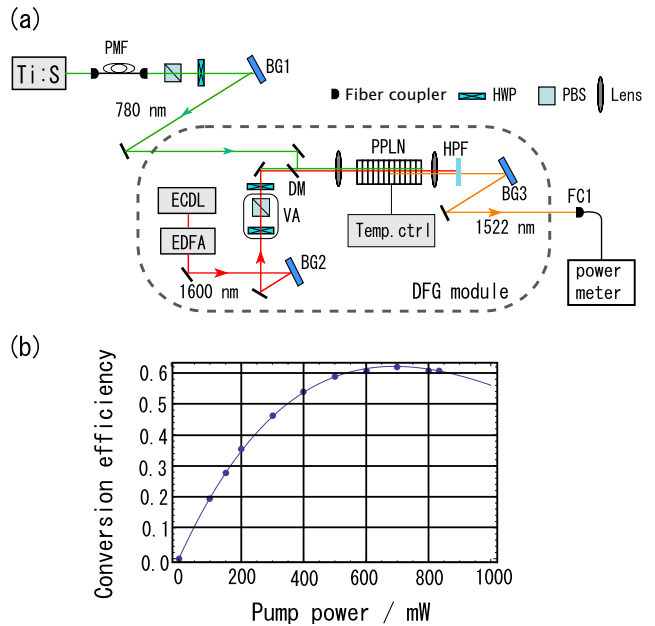}}
  \caption{
  Frequency down-conversion of coherent light pulse.
  (a) Experimental setup. 
  The signal light from the mode-locked Ti:sapphire~(Ti:S) 
  laser at 780 nm and the pump beam 
  from the external cavity diode laser~(ECDL) at 1600 nm 
  are combined at the dichroic mirror~(DM). 
  Then, they are coupled into the PPLN 
  for the frequency down-conversion to the 1522-nm wavelength. 
  The light at 1522 nm 
  is extracted using the high-pass filter~(HPF) and 
  the Bragg grating~(BG3) with a 1-nm bandwidth, 
  and coupled to the single-mode fiber. 
  (b) The conversion efficiency 
  depending on the power of the pump light coupled to the PPLN. 
  The maximum conversion efficiency $\approx 0.62$ was 
  obtained at 700-mW pump power. 
  A solid curve is fitted using a function $A \sin^2(\sqrt{BP\u{pump}})$ 
  expected from equation~(\ref{as}), 
  where $P\u{pump}$ is the pump power. 
  Parameters $A$ and $B$ are estimated as 
  0.62 and 3.6 $\U{(mW)^{-1}}$, respectively. 
  \label{fig:cl}}
 \end{center}
\end{figure}

As a preliminary experiment, 
we performed the frequency down-conversion 
of a classical light in a coherent state. 
Our experimental setup is shown in Fig.~\ref{fig:cl}~(a). 
We use a mode-locked Ti:sapphire~(Ti:S) laser 
as a signal light source. 
The center wavelength is 780 nm, 
the pulse width is 1.2 ps, and the repetition rate is 82 MHz. 
After passing through a polarization-maintaining fiber~(PMF), 
the signal beam is set to vertical~(V) polarization 
by a polarization beamsplitter~(PBS) and 
a half-wave plate~(HWP). 
The signal beam is diffracted by a Bragg grating~(BG1) 
which narrows the bandwidth to 0.2 nm. 
A seed laser beam from an external cavity diode laser~(ECDL) at 1600 nm 
with a linewidth of $\Delta f\equiv $150 kHz
is amplified by an Erbium-doped fiber amplifier~(EDFA), 
and then it is used as a pump light for the DFG. 
We use BG2 with the bandwidth of 1 nm 
for suppressing unnecessary frequency components of the pump light. 
The power of the pump light is adjusted 
by a variable attenuator~(VA) composed of a HWP and a PBS. 
The signal and pump lights are set to V polarization 
and combined at a dichroic mirror~(DM), 
and then focused on the PPLN crystal. 

The PPLN waveguide in our experiments 
consists of Zn-doped lithium niobate 
as the waveguide core and lithium tantalite 
as the cladding layer~\cite{nishikawa}. 
The waveguide is a ridged type with 8-$\mu$m wide, 
and its length is 20 mm. 
The periodically-poled structure has an 18-$\mu$m period 
designed for the type-0 quasi-phase matching. 
The best conversion efficiency was obtained 
at the temperature of 50$^\circ$C. 
The acceptable phase-matching bandwidth 
for the 780-nm signal light of the 20-mm long crystal 
is calculated to be about 0.3 nm. 
 
After passing through the PPLN waveguide, 
the strong pump light is diminished 
by a high-pass filter~(HPF) 
and the converted light at 1522 nm 
is extracted by BG3 with a bandwidth of 1 nm. 
Finally the beam is coupled into a single-mode fiber 
by a fiber coupler~(FC1) and connected to a power meter. 
The photon conversion efficiency of the PPLN was observed 
as shown in Fig.~\ref{fig:cl}~(b). 
The maximum conversion efficiency was 
$\approx 0.62$ at 700-mW pump power coupled to the PPLN waveguide. 
The transmittance of the optical components 
from the end of the PPLN 
to the power meter was estimated to be $\approx 0.62$. 
This gives the overall conversion efficiency 
from the 780-nm photon coupled in the PPLN 
to the 1522-nm photon coupled to the fiber as $\approx 0.39$. 
The conversion efficiency will be improved 
when the linewidth of the signal light 
becomes much narrower than the acceptable linewidth of the PPLN waveguide. 
When we used a 780-nm cw light from an ECDL with a 3-MHz linewidth, 
we observed the maximum conversion efficiency of 0.71. 

\begin{figure}
 \begin{center}
 \scalebox{1}{\includegraphics{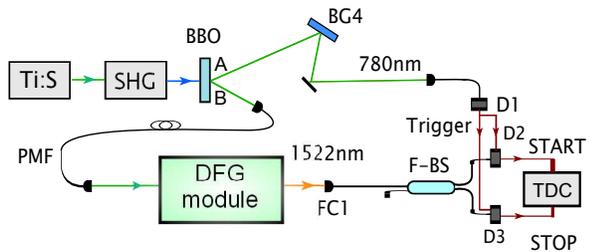}}
  \caption{
  Frequency down-conversion of heralded single photons at 780 nm. 
  The light pulse from the Ti:S laser is frequency doubled 
  to the wavelength of 390 nm by second harmonic generation~(SHG), 
  and then pumps 
  a Type-I phase-matched 1.5mm-thick BBO crystal 
  to prepare a photon pair A and B at 780 nm 
  by spontaneous parametric down-conversion. 
  The photon in mode A is detected by a silicon avalanche photodiode 
  detector D1 connected to a single-mode fiber for 
  the preparation of the heralded single photon in mode B. 
  We observe the second-order intensity correlation function $g^{(2)}_0$ 
  of the light converted to 1522 nm 
  using the Hanbury-Brown and Twiss setup. 
  \label{fig:ql}}
 \end{center}
\end{figure}

The preservation of the photon statistics of the signal light 
after the frequency conversion 
is the first step 
toward the demonstration of our quantum interface. 
Non-classical behavior of the photon statistics 
of the pulsed light 
appears in the second-order intensity correlation function 
between the $n$th neighbor pulses defined by 
\begin{eqnarray}
g^{(2)}_n\equiv 
\frac{\displaystyle
\expect{:\int\hat{I}(t)\U{d}t \int\hat{I}(t+nT)\U{d}t :}}
{\displaystyle
\expect{\int\hat{I}(t)\U{d}t}\expect{\int\hat{I}(t+nT)\U{d}t}}, 
\end{eqnarray}
where $\hat{I}(t)$ is an intensity operator at time $t$, 
$T$ is the time interval of the pulses, and 
the integrations are over a range of the single-pulse duration. 
The product of the operators is 
in normal order and time order~\cite{g2}. 
While $g^{(2)}_0\geq 1$ is always satisfied 
in the classical wave theory, 
the quantization of light 
allows the possibility that $g^{(2)}_0$ is smaller than 1, 
with the smallest value $g^{(2)}_0=0$ 
for an ideal single-photon pulse. 
The experimental setup for 
frequency down-conversion of a single photon at 780 nm 
followed by correlation measurement on the converted light 
is shown in Fig.~\ref{fig:ql}. 
A photon pair A and B are generated 
by spontaneous parametric down-conversion~(SPDC) 
at a $\beta$-barium borate~(BBO) crystal, 
and the presence of a single photon in mode B is heralded 
by detection of a photon in mode A by a photon detector D1. 
The spectral filtering of the photon in mode A is 
performed by BG4 with a bandwidth of 0.2 nm. 
The wavelength of the heralded single photon is converted 
to 1522 nm by the DFG module described in Fig.~\ref{fig:cl}~(a). 
The converted pulse is coupled to FC1, and 
its second-order intensity correlation $g^{(2)}_0$ is measured 
by using a Hanbury-Brown and Twiss setup~\cite{g21, g22}, 
which consists of a fiber-optic beamsplitter (F-BS) and 
photon detectors D2 and D3. 
D2 and D3 are InGaAs/InP avalanche photodiodes 
gated by trigger signals from D1. 
To measure the temporal difference 
between the detections at D2 and D3 with high resolution, 
signals from D2 and D3 are input 
to a time-to-digital converter~(TDC) 
as a start and a stop signal of the clock, respectively. 
By recording the trigger counts $N\u{trig}$ from D1, 
the start counts $N\u{start}$ from D2, 
the stop counts $N\u{stop}$ from D3, 
and coincidence counts $N\u{coinc}$ within 1 ns from the TDC, 
we determine $g^{(2)}_0$ by 
\begin{eqnarray}
g^{(2)}_0=\frac{N\u{trig}N\u{coinc} }{N\u{start}N\u{stop}}. 
\end{eqnarray}

The measured $g^{(2)}_0$ of the light converted 
from the heralded single photon at 780 nm 
was $0.17\pm 0.04$, 
which is much smaller than one, and indicates 
sub-Poissonian photon statistics of the converted light. 
This result clearly shows 
that the non-classical property of the light 
survived after the frequency down-conversion. 
The nonzero value of $g^{(2)}_0$ is mainly 
caused by optical noises from the PPLN. 
They are linearly increasing with the pump power 
when the signal light is turned off, 
which may suggest that they are caused by Raman scattering~\cite{Raman}. 

\begin{figure}
 \begin{center}
 \scalebox{1}{\includegraphics{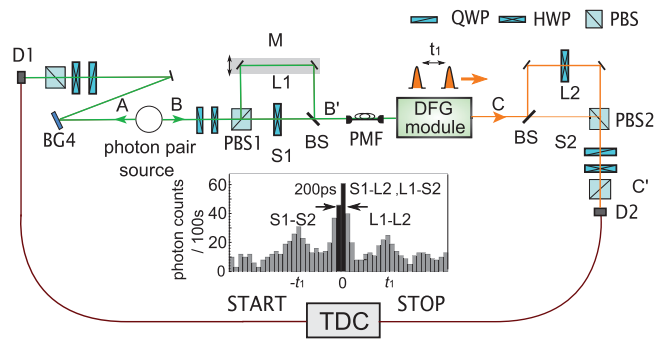}}
  \caption{
  Frequency down-conversion of 
  one halves of polarization-entangled photon pairs. 
  At the photon pair source, 
  the light pulse from the Ti:S laser is frequency doubled by SHG, 
  and pumps a pair of BBO crystals 
  to prepare 
  the entangled photon pair $\ket{\phi^+}\u{AB}$ through SPDC~\cite{DFS}. 
  The frequency down-conversion of photon B is performed 
  after encoding a polarization qubit into 
  a time-bin qubit using an unbalanced MZI, 
  then decoding it back to the polarization qubit 
  after frequency down-conversion. 
  The histogram shows the number of coincidence events 
  with various delay between the detectors D1 and D2, 
  which were recorded by the TDC. 
  The measurement bases of photons in modes A and $\U{C'}$ 
  are set as +45$^\circ$ polarization. 
  The central peak shows the events 
  where the photon has passed through S1-L2 or L1-S2, 
  and the two peaks separated from the central peak 
  by $t\u{1}$ correspond to the cases 
  where the photon has passed through S1-S2 or L1-L2. 
  We accept the events in 200-ps time window 
  around the central peak as the successful events. 
  \label{fig:bellpair}}
 \end{center}
\end{figure}
Finally, we performed the frequency down-conversion 
of one halves of 
polarization-entangled photon pairs at 780 nm. 
The experimental setup is shown in Fig.~\ref{fig:bellpair}. 
A photon pair source prepares 
a polarization-entangled photon pair A and B 
described as 
$\ket{\phi^+}\u{AB}\equiv 
(\ket{\H\H}\u{AB}+\ket{\V\V}\u{AB})/\sqrt{2}$. 
The spectral filtering of photon A is performed by BG4. 
Photon B goes to an unbalanced Mach-Zehnder interferometer~(MZI). 
At the MZI, PBS1 separates H and V polarization 
into a short path~(S1) and a long path~(L1), respectively. 
After horizontal~(H) polarization is flipped to V polarization 
by a HWP inserted into S1, 
modes S1 and L1 are mixed by a BS. 
This transforms the polarization qubit in mode B 
to a time-bin qubit in mode $\U{B'}$, and 
the initial two-photon state $\ket{\phi^+}\u{AB}$ becomes 
a polarization and time-bin entangled state 
\begin{eqnarray}
\ket{\psi}\u{AB'} \equiv 
\frac{1}{\sqrt{2}}(\ket{\H}\u{A}\ket{\U{S1}}\u{B'}
+\ket{\V}\u{A}\ket{\U{L1}}\u{B'}), 
\label{path}
\end{eqnarray}
where $\ket{\U{S1}}$ and $\ket{\U{L1}}$ 
represent the states of the V-polarized photons 
passing through S1 and L1, respectively. 
After passing through a PMF, 
the photon in mode $\U{B'}$ goes to the DFG module. 

The light down-converted by the DFG process 
inherits the phase $-\phi$ of the pump light 
as described in equation~(\ref{as}). 
Then, through the frequency down-conversion, 
the former and the latter term in equation~(\ref{path}) 
receive phase shifts 
$-\phi(t)$ and $-\phi(t+t\u{1})$ from the pump light, respectively. 
In our experiment, since 
$t\u{1}$ is about 1 ns and $\Delta f$ is 150 kHz, 
the coherence time $1/(2\pi \Delta f)\approx 1\mu$s 
of the pump light is much longer than $t\u{1}$, 
which ensures that $\phi(t)=\phi(t+t\u{1})$ is satisfied. 
As a result, we should obtain the state $\ket{\psi}\u{AC}$ 
for the 780-nm photon 
and the 1522-nm photon in mode C. 

The down-converted photon in mode C 
is fed to the second unbalanced MZI. 
A BS in the MZI divides the converted photon 
into a short path~(S2) and a long path~(L2). 
A HWP inserted in L2 flips the polarization 
from V to H. 
Two modes S2 and L2 are recombined by PBS2. 
This transforms the degree of freedom 
from time-bin to polarization. 
After the MZI, the photon in mode $\U{C'}$ 
goes to detector D2 which is gated by trigger counts from D1. 
The path lengths S1-L2 and L1-S2 are 
adjusted by the mirrors M 
on a motorized and piezo-driven stage. 
When we post-select the events 
where the photon has passed through S1-L2 or L1-S2, 
we should obtain the polarization-entangled state 
$\ket{\phi^+}\u{AC'}$ for the two photons in modes A and $\U{C'}$. 
In our experiment, we accept such events in 200-ps time window 
shown in the histogram in the inset of Fig.~\ref{fig:bellpair}. 

\begin{figure}[t]
 \begin{center}
  \scalebox{1}{\includegraphics{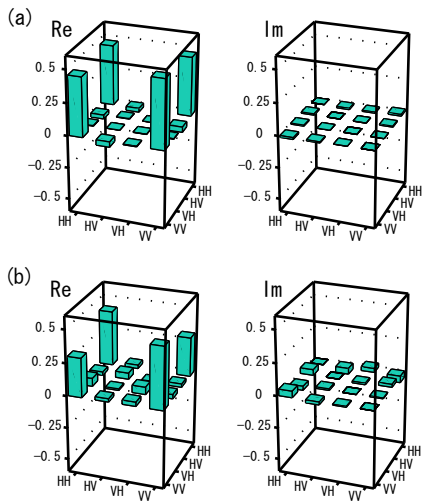}}
  \caption{
  The reconstructed density matrices.
  (a) The initial entangled photon pair $\rho\u{AB}$ 
  prepared by the photon pair source. 
  (b) The two-photon state $\rho\u{AC'}$ 
  between the 780-nm photon in mode A and 
  the 1522-nm photon in mode $\U{C'}$. 
  The overall detection rate of each photon pair 
  $\rho\u{AB}$ and $\rho\u{AC'}$ 
  is 492 Hz and 2.4 Hz, respectively. 
  \label{fig:matrix}}
 \end{center}
\end{figure}
We first performed quantum state tomography of 
780-nm photon pairs in modes A and B 
from the photon pair source. 
The required polarization correlations have been measured 
by rotating a quarter-wave plate~(QWP) and a HWP in mode A, 
and a QWP and a HWP before PBS1 in mode B~\cite{MQ}. 
The PMF before the DFG module was connected to a silicon avalanche photodiode. 
The observed density operator $\rho\u{AB}$ of the two-photon state 
is shown in Fig.~\ref{fig:matrix}~(a). 
The iterative maximum likelihood method was used 
for the reconstruction of $\rho\u{AB}$~\cite{MLE1, MLE2}. 
The observed fidelity of $\rho\u{AB}$ 
to $\ket{\phi^+}\u{AB}$ was $0.95\pm 0.01$, 
which implies that the photon pair source prepared 
a highly entangled 780-nm photon pair. 
Next, we performed quantum state tomography 
between the photon in mode A and the converted photon in mode $\U{C'}$. 
The reconstructed density operator $\rho\u{AC'}$ 
of the state is shown in Fig.~\ref{fig:matrix}~(b). 
We calculated the fidelity and 
entanglement of formation~\cite{wooters} 
from reconstructed $\rho\u{AC'}$ 
as $0.75\pm 0.06$ and $0.36\pm 0.13$, respectively. 
The observed value of the fidelity is greater than 
$1/\sqrt{2}\approx 0.707$, which implies that 
the entangled state $\rho\u{AC'}$ 
can violate the Clauser-Horne-Shimony-Holt~(CHSH) inequality. 
We note that when we subtract background noises 
whose rate is about 0.2 Hz in each basis 
from the observed coincidence counts, 
the fidelity of the output state becomes 
$0.95$, 
which shows that the degradation 
is mainly caused by the optical noises from the PPLN. 

In conclusion, we have demonstrated frequency down-conversion 
of non-classical light at 780 nm 
to a telecommunication wavelength of 1522 nm 
by using the DFG process in the PPLN waveguide. 
Our quantum interface has 
down-converted the frequency of one half 
of the visible entangled photon pair 
to the telecommunication wavelength 
while preserving entanglement. 
The DFG process with the quasi-phase-matching techniques 
allows us to convert a wide range of visible wavelengths 
to telecommunication ones with a wide acceptable bandwidth. 
We believe such a quantum interface will be useful 
for long distance quantum communication 
based on various visible photon emitters 
including atomic quantum memories. 

We thank Toshiyuki Tashima and Hirokazu Kobayashi 
for helpful discussions. 
This work was supported by the Funding Program for 
World-Leading Innovative R \& D on Science and 
Technology (FIRST), MEXT Grant-in-Aid for Scientific 
Research on Innovative Areas 20104003 and 21102008, 
MEXT Grant-in-Aid for Young scientists(A) 23684035, 
and the MEXT Global COE Program.


\begin{thebibliography}{999}

\bibitem{E1} 
D. N. Matsukevich and A. Kuzmich, 
Science {\bf 306}, 663 (2004). 

\bibitem{E2} 
W. Rosenfeld {\it et al.}, 
\prl {\bf 101}, 260403 (2008). 

\bibitem{E4} 
S. Olmschenk {\it et al.}, 
Science {\bf 323}, 486 (2009). 

\bibitem{E5} 
E. Togan {\it et al.}, 
Nature {\bf 466}, 730 (2010). 

\bibitem{E6} 
T. M\"uller {\it et al. }
cond-mat/arXiv:1101.4911 (2011). 

\bibitem{R1} 
H. -J. Briegel {\it et al.}, 
\prl {\bf 81}, 5932 (1998). 

\bibitem{R2} 
L. M. Duan {\it et al.}, 
Nature (London) {\bf 414}, 413 (2001). 

\bibitem{CZ1}
J. I. Cirac and P. Zoller, 
Nature (London) {\bf 404}, 579 (2000). 

\bibitem{Ex1} 
Z. Zhao {\it et al.}, 
\prl {\bf 90}, 207901 (2003). 

\bibitem{Ex2} 
P. Walther {\it et al.}, 
Nature (London) {\bf 434}, 169 (2005). 

\bibitem{communication} 
N. Gisin and R. Thew, 
Nature Photonics {\bf 1}, 165 (2007). 

\bibitem{conv} 
P. Kumar, 
Opt. Lett. {\bf 15}, 1476 (1990). 

\bibitem{PPLN1} 
S. Tanzilli {\it et al.},
Nature (London) {\bf 437}, 116 (2005). 

\bibitem{Langrock} 
C. Langrock, {\it et al.}, 
Opt. Lett. {\bf 30}, 1725 (2005). 

\bibitem{four1} 
A. G. Radnaev {\it et al.}, 
Nature Physics {\bf 6}, 894 (2010). 

\bibitem{four2} 
Y. O. Dudin {\it et al.}, 
\prl {\bf 105}, 260502 (2010). 

\bibitem{DFG1} 
H. Takesue, 
\pra {\bf 82}, 013833 (2010). 

\bibitem{DFG2} 
N. Curtz {\it et al.}, 
Opt. Express {\bf 18}, 22099 (2010).

\bibitem{nishikawa}  
T. Nishikawa {\it et al.}, 
Opt. Express {\bf 17}, 17792 (2009). 

\bibitem{g2} 
M. Koashi {\it et al.}, 
\prl {\bf 71}, 1164 (1993).

\bibitem{g21} 
R. H. Brown and R. Q. Twiss, 
Nature {\bf 178}, 481 (1956). 

\bibitem{g22} 
M. Beck, 
J. Opt. Soc. Am. B. {\bf 24}, 2972 (2007). 

\bibitem{Raman} 
N. A. Peters {\it et al.},  
New J. Phys. {\bf 11}, 045012 (2009).

\bibitem{DFS} 
T. Yamamoto {\it et al.}, 
Nature Photonics {\bf 2}, 488 (2008). 

\bibitem{MQ} 
D. F. V. James {\it et al.}, 
Measurement of qubits. 
\pra {\bf 64}, 052312 (2001). 

\bibitem{MLE1} 
J. $\U{\check{R}eh\acute{a}\check{c}ek}$, and Z. Hradil, 
\pra {\bf 75}, 042108 (2007). 

\bibitem{MLE2} 
T. Tashima {\it et al.}, 
\prl {\bf 102}, 130502 (2009). 

\bibitem{wooters} 
W. K. Wootters, 
\prl {\bf 80}, 2245 (1998).

\end{thebibliography}
\end{document}